# Ehrenfest's adiabatic hypothesis in Bohr's quantum theory

Enric Pérez* and Blai Pié Valls**

**Abstract:** It's widely known that Paul Ehrenfest formulated and applied his adiabatic hypothesis in the early 1910s. Niels Bohr, in his first attempt to construct a quantum theory in 1916, used it for fundamental purposes in a paper which eventually did not reach the press. He decided not to publish it after having received the new results by Sommerfeld in Munich. Two years later, Bohr published "On the quantum theory of line-spectra." There, the adiabatic hypothesis played an important role, although it appeared with another name: the *principle of mechanical transformability*. In the subsequent variations of his theory, Bohr never suppressed this principle completely.

We discuss the role of Ehrenfest's principle in the works of Bohr, paying special attention to its relation to the correspondence principle. We will also consider how Ehrenfest faced Bohr's uses of his more celebrated contribution to quantum theory, as well as his own participation in the spreading of Bohr's ideas.

**Key words:** Adiabatic Principle, Correspondence Principle, Paul Ehrenfest, Niels Bohr.

## 1. Introduction

Ninety years ago an issue of *Die Naturwissenschaften* was released to commemorate the tenth anniversary of Bohr's atom (Figure 1). Ehrenfest contributed a paper on the adiabatic principle, and the contributions by Kramers, and by Bohr himself (in fact, his Nobel lecture) also mentioned Ehrenfest's hypothesis. In a similar vein, it is not uncommon to find references to the adiabatic principle when considering Bohr's quantum theory from a historiographical point of view, as one of its two pillars, next to the correspondence principle.[1] Yet, the disproportion in the interest provoked and the treatment given to the two principles is significant. In fact, in the commemorative issue of 1923, besides the aforementioned allusions, the adiabatic principle hardly appears.

In this paper we want to deal with the role of the adiabatic principle in Bohr's thoughts in the years of the old quantum theory, mainly between 1918 and 1923, and gauge its relative importance. To do this, in addition to analyzing the works of Bohr, we will sketch how Ehrenfest reacted to the slow transformation of his hypothesis.

---

* Departament de Física Fonamental, Universitat de Barcelona. E-mail: enperez@ub.edu.

** Departament de Física Fonamental, Universitat de Barcelona. E-mail: blai.pie.valls@ub.edu.

[1] For instance, Jammer (1966), Darrigol (1992), Kragh (2012).

## 2. Ehrenfest adiabatic hypothesis

In 1913, Ehrenfest transformed infinitely slowly Planck's oscillators into diatomic molecules to find the quantization of their angular momentum.[2] In the same year, he drafted his first work exclusively devoted to the adiabatic hypothesis.[3] The paper was poorly written, and published only in the *Proceedings of the Amsterdam Academy*. The basic idea underlying his treatment was that allowed quantum motions transform into allowed quantum motions during adiabatic transformations.

Either because of the advent of war or because of the schematic character of the article, the proposal by Ehrenfest went nearly unnoticed and had almost no influence on the subsequent attempts to formulate quantization rules. Therefore, after checking the compatibility between his adiabatic hypothesis and Sommerfeld's results of 1915/1916, Ehrenfest quickly wrote a more elaborate article, which also included the results of another work of 1914 by himself in which he had found a necessary condition to be satisfied by statistical weights in order to maintain the validity of Boltzmann's principle.[4]

In a letter to Sommerfeld he drew attention to his unknown contribution. In this letter, Ehrenfest expressed, in his characteristic way, his regret that the work from Munich had contributed to the success of Bohr's model:[5]

> Understandably, your work and the subsequent success of Epstein provided me and my friends very great pleasure. Even though I think it is appalling that this success will help the provisional but still so cannibalistic Bohr model to obtain new triumphs - I warmly wish the Munich Physics further success on this path!

Certainly, Ehrenfest initially considered Bohr's ideas loathsome. So he confessed to his Ukrainian friend Abraham Ioffé:[6]

> Bohr's work 'quantum mechanical consequences of Balmer's law' [sic] bothers me: if

---

[2] Ehrenfest (1913a). On Ehrenfest's adiabatic hypothesis see Klein (1970), Navarro & Pérez (2004), Navarro & Pérez (2006), Pérez (2009).

[3] Ehrenfest (1913b).

[4] Ehrenfest (1914, 1916). With 'Boltzmann's principle' we are referring to the known relation between entropy and probability postulated by Boltzmann.

[5] Ehrenfest to Sommerfeld, April/May 1916. Sommerfeld (2000), pp. 555-557.

[6] Ehrenfest to Ioffé, 28 August 1913. Moskovchenko and Frenkel (1990), p. 122.

the Balmer formula can be obtained in this way, I must throw all the physics in the trash (and myself...)

Needless to say, this assessment was not reciprocal at all. In his response to Ehrenfest, Sommerfeld informed him that Bohr had already told him about the adiabatic hypothesis in a very approving way. Bohr had written to Sommerfeld that he "... had made considerable use of Ehrenfest's idea about adiabatic transformations which seems to me very important and fundamental..."[7] This was March 1916. Let us now go back and see when Bohr used the adiabatic hypothesis for the first time.

## 3. Bohr's trilogy and his quantum theory of periodic systems (1913-1916)

In the trilogy by Bohr, Ehrenfest's adiabatic hypothesis did not appear. In fact, as for dates, Bohr could not have known the first version of the hypothesis before sending to print his first installment of "On the Constitution of Atoms and Molecules." But neither is it mentioned in the next two. What we do find in the second and third installments are slow imaginary transformations of certain annular configurations and even the process of formation of the hydrogen molecule through an approach of two hydrogen atoms, "so slow that the dynamical equilibrium of the electrons for every position of the nuclei is the same as if the latter were at rest."[8] Nevertheless, Bohr neither mentioned adiabatic invariants nor explicitly discussed the validity of mechanics in those kind of transformations, as he would do in later papers.

By the spring of 1916, he had written down a theory that would group and base quantization for periodic systems. But when the paper was about to be published in *Philosophical Magazine*, Bohr received the new developments on multiperiodic systems by Sommerfeld,[9] and decided to retract it. He only published "On the Application of the Quantum Theory to Periodic Systems"[10] in 1921, when it was a kind of historical curiosity, and he did it because of the insistence by Ehrenfest, among others, who encouraged him by saying that its publication could help readers to follow his thought processes, so difficult to understand "even for an Einstein."[11]

---

[7] Bohr to Sommerfeld, 19 March 1916. BCW3 1981, p. 604.

[8] Bohr (1913), pp. 481-482, p. 868.

[9] For further reading, see the contribution by Michael Eckert to this volume.

[10] Bohr (1916). On this paper see Darrigol (1992), pp. 93-98.

[11] Ehrenfest to Bohr, 30 January 1920. AHQP, microfilm AHQP/BSC-2.

Bohr's theory of 1916 was based on the assumptions that there exist stationary states in atomic systems and that "any change of the energy of the system including absorption and emission of electromagnetic radiation must take place by a transition between two such states."[12] Among the conditions they have to fulfill we find:

$$\frac{\bar{T}}{\omega} = \oint T dt = \frac{hn}{2},\qquad(1)$$

($\bar{T}$ is the average kinetic energy, $\omega$ the frequency of the motion, $t$ the time, $h$ Planck's constant, and $n$ a whole number). According to Bohr, the "possibility of a consistent theory based on this assumption is given by:"

$$\delta W = 2\omega\delta\left(\frac{\bar{T}}{\omega}\right).\qquad(2)$$

Which is the difference of the total energy for two neighbouring periodic motions of the same system. In other words, invariant (1)—whose validity Bohr extends to relativistic systems—is the quantity to be quantized; there is emission and absorption of radiation only in transitions between stationary states; (2) implies that the energy of a periodic system is completely determined by the value of the adiabatic invariant

$$\frac{\bar{T}}{\omega}.$$

Here Bohr quotes Ehrenfest, who had pointed out the "great importance in the Quantum theory of this invariant character of [(1)]." And he adds that Ehrenfest's idea:

> … allows us by varying the external conditions to obtain a continuous transformation through possible states from a stationary state of any periodic system to the state corresponding with the same value of *n* of any other such system containing the same number of moving particles.

---

[12] Bohr (1916), p. 434.

Bohr justifies this invariance by arguing that in the cases when the external field is established slowly and at a uniform rate, it can be "considered as an inherent part of the system," and then the internal motion of the whole system obeys ordinary mechanics.[13]

**4. "On the quantum theory of line-spectra" (1918)**

The updating of his theory took Bohr two years.[14] In the first part of the new (and published) version of his quantum theory, of 1918, he dealt with multiperiodic systems. There, Bohr presented the work of Ehrenfest as one of the great advances obtained recently in the quantum theory along with Einstein's transition probabilities and the magnificent developments by the Munich school. In "On the quantum theory of line-spectra," Bohr quotes practically all the papers by Ehrenfest related to the adiabatic issue and formulates the most complete version of his hypothesis. In part I he even quotes the papers where Burgers extended the validity of the results of Ehrenfest, and in part II Burgers' dissertation "Het Atommodel van Rutherford-Bohr," appeared in 1918.[15]

Here Ehrenfest's idea has a much more fundamental role than in 1916. Now, Bohr establishes a principle according to which mechanics still applies in continuous transformations. That guarantees the stability of stationary states. Moreover, he includes the statistical implications to properly generalize the meaning of Boltzmann's principle within the quantum theory. Bohr coins the term *principle of mechanical transformability*, in order to take distance from the thermodynamic reminiscences which, according to Bohr, Ehrenfest's hypothesis had.

Thus, in a variation of the external conditions, systems usually readjust in a non-mechanical way. If, however,[16]

> … [a slow] variation is performed at a constant or very slowly changing rate, the forces
> to which the particles of the system will be exposed will not differ at any moment
> from those to which they would be exposed if we imagine that the external forces arise from a number of slowly moving additional particles which together with the original system form a system in a stationary state.

---

[13] Bohr (1916), p. 436.

[14] Bohr (1918a).

[15] Bohr (1918a), p. 17, Bohr (1918b), p. 93, footnote. Bohr writes that Burgers "has given a very interesting general survey of the applications of the quantum theory to the problem of the constitution of atoms, and has in this connection entered upon several questions discussed in the present paper".

[16] Bohr (1918a), p. 8.

For simply periodic motions Bohr derives the adiabatic invariance of:

$$I = \int_0^\sigma \sum_1^s p_k \dot{q}_k dt$$

($\sigma$ is the period of motion, $s$ the number of degrees of freedom, and $p_k$ and $q_k$ phase coordinates). The quantization is given by the formula:

$$I = nh.$$

And the natural generalization for the multiperiodic case is:

$$I_k = n_k h, \tag{3}$$

with

$$I_k = \int p_k dq_k$$

($k=1\cdots r$, where $r$ is the degree of periodicity, and the integral is extended over each partial period) for each periodic component in which the motion can be separated, according to the theory of Hamilton-Jacobi. In degenerate cases, where the motion of the phase point does not cover densely a $s$-dimensional extension, the separation of variables is ambiguous, and so too is the quantization given by (3). Accordingly, for slow transformations, passing through degenerate motions constitutes a singularity: there the system has to adapt in a non-mechanical way, as in fast changes. Bohr notes that this is so because of the appearance/disappearance of new vibrations: the transformation cannot be slow any more with respect to the period of the new vibration, which is very long near the point of degeneracy. Bohr takes advantage of this peculiar fact to devise transformations that connect in a continuous way different stationary states of the same system; this is what he calls "cyclic transformations."

As for the statistical treatment, we read:[17]

> In examining the necessary conditions for the explanation of the second law of thermodynamics Ehrenfest has deduced a certain general condition as regards the variation of the a-priori probability corresponding to a small change of the external conditions from which it follows that the a-priori probability of a given stationary state of an atomic system must remain unaltered during a continuous transformation, except in special cases in which the values of the energy in some of the stationary states will tend to coincide during the transformation. In this result we possess, as we shall see, a rational basis for the determination of the a-priori probability of different stationary states of a given atomic system.

In other words, the quantization of adiabatic invariants meets the preconditions for the applicability of Boltzmann's principle.

**5. Ehrenfest's first reaction**

Bohr sent the first part of "On the quantum theory of line-spectra" to Ehrenfest in May 1918, along with a letter explaining the change in terminology for his hypothesis, now a principle. As Bohr explained, in the published memoir he had not been explicit enough in this respect for the sake of brevity:[18]

> As you will see the considerations are to a large extent based on your important principle of 'adiabatic invariance.' As far as I understand, however, I consider the problem from a point of view which differs somewhat from yours, and I have therefore not made use of the same terminology as in your original papers. In my opinion the condition of the continuous transformability of motion in the stationary states may be considered as a direct consequence of the necessary stability of these states and to my eyes the main problem consists therefore in the justification of the application of ordinary 'mechanics' in calculating the effect of a continuous transformation of the system.

---

[17] Bohr (1918a), pp. 9-10. See Darrigol (1992), pp. 132-137.

[18] Bohr to Ehrenfest, 5 May 1918. AHQP, microfilm AHQP/EHR-17, Section 5.

It is interesting to read Ehrenfest's response.[19] At that time, he was "far away from physics," as well as "suffering an attack of jaundice, depressed by the interminable war and dissatisfied with his own work."[20] In fact, he did not answer Bohr's letter until three months later, and even then still without having read carefully the work of Bohr.

Ehrenfest completely agreed with Bohr that, were the "*Transformationsprinzip*" right, it would be more fundamental than thermodynamics. However, he added some interesting observations on why, despite appearances to the contrary, he claimed not to have given his own hypothesis the thermodynamical meaning Bohr took him to have had in mind. First of all, his idea was inspired by Helmholtz's and Boltzmann's considerations on (mechanical) monocycles; Ehrenfest thought that the expression "principle of mechanical transformability" should somehow be restricted, because it suggests that any motion could be transformed mechanically into any other motion, and in that case invariants become meaningless. Moreover, adiabatic transformations of statistical weights in a system in thermodynamical equilibrium do not necessarily lead to a new state of equilibrium, because the transformation is performed in phase space, not in the space of macroscopic variables. In this sense, his previous uses of the term "adiabatic" departed clearly from thermodynamics.

In this first letter Ehrenfest informed Bohr about the celebration, the following Easter, of the Dutch Congress of Natural and Medical Sciences in Leyden and invited him to attend. From then on, correspondence between the two physicists shows how their friendship grew. They met for the first time in Leyden, at the aforementioned conference.

**6. Bohr's second fundamental postulates**

By mid-1922 Bohr had finished a theoretical basis for his second atomic theory, which strongly departed from the ring model of 1913.[21] He presented it in his Wolfskehl lectures in Göttingen in June, and it appeared as a paper in January 1923 under the title: "On the Application of the Quantum Theory to Atomic Structure." Only the first part was published: "The Fundamental Postulates."

---

[19] Ehrenfest to Bohr, 14 July 1918. AHQP, microfilm AHQP/BSC-2, Section 1.

[20] Klein (2010), p. 308.

[21] Bohr (1924). For Bohr's second atomic theory, see Kragh (2012), pp. 272-297.

There, after stating the old assumptions related to the existence of stationary states, Bohr writes the formula which provides "the displacement of a particle in a given direction:"

$$\xi = \sum C_{\tau_1 \cdots \tau_u} \cos 2\pi \big([\tau_1 \omega_1 + \cdots + \tau_u \omega_u]t + \gamma_{\tau_1 \cdots \tau_u}\big),$$

where $\omega_1, \cdots, \omega_u$ are the fundamental frequencies and $u$ is the degree of periodicity. He adds that:[22]

> The summation is to be extended to all integral values of the numbers $\tau_1, \cdots, \tau_u$. The uniqueness of the solution is conditioned by the fact that among the quantities $\omega_1, \cdots, \omega_u$ there exist no relations of the form:
>
> $$m_1 \omega_1 + \cdots + m_u \omega_u = 0, \qquad [(4)]$$
>
> where $m_1, \cdots, m_u$ are a series of whole numbers.

This way of introducing his new theory nicely captures the evolution of Bohr's thoughts in the years after 1918. Although this idea appeared in Part II of "On the quantum theory of line-spectra," it took a more central role in the subsequent years. From 1920 on, the importance of the correspondence principle in the foundations of his theory became stronger.[23] Relation (4) is precisely the condition that a non-degenerate movement must fulfill. Now, it is closely tied with the correspondence principle and, in fact, explains, along with the adiabatic principle, why decreasing the degree of degeneracy is equivalent to the appearance of new frequencies:

> The addition of further conditions, if the degree of periodicity increases under the influence of external forces, appears too in a very simple light. We can, in fact, regard these conditions as an immediate demand for a correspondence between the new, slow harmonic vibrations appearing in the secular perturbations and processes of transition, for which the quantum numbers already appearing in the undisturbed motion are not changed, but only the new quantum numbers, appearing in the additional conditions.

---

[22] Bohr (1924), p. 4.

[23] For further reading, see the contribution by Martin Jähnert to this volume.

Therefore, a guideline originally devised to analyze and characterize multiperiodic systems turned into a crucial tool to formulate the correspondence principle. Relation (4) also establishes that the degree of periodicity is the significant parameter of a system, not the number of dimensions:[24]

> The assertion that the number of quantum conditions… is exactly equal to the degree of periodicity, becomes a necessary demand for attaining an unambiguous correspondence between the various types of transitions and the harmonic components appearing in the motion.

Bohr recovers the terminology 'adiabatic principle' without justifying the new change, maybe because he now refers mostly to the validity of electrodynamics, not to mechanics:[25]

> We may say that the Adiabatic Principle ensures the stability of the stationary states in the region in which we might on the whole expect that this stability can be discussed on the basis of the ordinary electrodynamic laws.

Mechanics was definitely losing ground: ordinary mechanics does not apply to the motion of any stationary state. The correspondence principle contributed strongly to emphasize the new role of electrodynamics. However, Bohr kept assuming the existence of quantum numbers even in complex systems, despite the fact that equations of motion cannot be solved:[26]

> In the fixation of these quantum numbers considerations which rest on the Adiabatic Principle, as well as on the Correspondence Principle discussed in the next chapter, play an important role. The demand for the presence of sharp, stable, stationary states can be referred to, in the language of the quantum theory, as a general principle *of the existence and permanence of the quantum numbers*.

---

[24] Bohr (1924), p. 25.

[25] Bohr (1924), p. 14.

[26] Bohr (1924), p. 16.

That is, some transformations do not change the quantum number, even if they are no longer associated with the validity of mechanics. This use, already dissociated from mechanics, was meant to the construction of some atomic and molecular models.

In sum, we conjecture that the less mechanistic character of his theory made Bohr come back to the original denomination of Ehrenfest's principle. The adiabatic transformation of statistical weights, originally subordinated to the validity of mechanics, remained as the essence of Ehrenfest's hypothesis. Hence, it became more a statistical than a mechanical principle.

## 7. Ehrenfest on the correspondence principle

Meanwhile, after their first personal contact, Ehrenfest had plunged into Bohr's atomic theory, publishing papers devoted to polishing and developing the results of his colleague, some of them with ingenious applications of the correspondence principle.[27] The most telling episode of this subordination of Ehrenfest's research interests to those of Bohr occurred during the 3rd Solvay Conference in the spring of 1921, which Bohr could not attend for health reasons. On his behalf, Ehrenfest presented a paper in which he analyzed the implications and assumptions underlying the correspondence principle by Bohr and Kramers;[28] he only made a slight allusion to the adiabatic principle when dispersion was discussed after his communication.

The next Christmas, in December 1921, Ehrenfest spent three weeks in Copenhagen with his daughter Tatiana (Pavlovna); he gave a couple of talks, one of them on the "mystery of energy quanta."[29] In one of the postcards he wrote to Bohr during a tour around Scandinavia, he asked what Bohr meant with "a certain extension of Adiabatic Hypothesis."[30] Probably, Bohr was then working on the theory we have already outlined, and he had been discussing it vividly with Ehrenfest. Indeed, according to Bohr, the conversations they had on that visit greatly influenced his thoughts. Some months later, referring to his new paper, he wrote to Ehrenfest:[31]

---

[27] Ehrenfest and Breit (1922).

[28] Ehrenfest (1922).

[29] Ehrenfest to Bohr, 11 November 1921. AHQP, microfilm AHQP/BSC-2, Section 2.

[30] Ehrenfest to Bohr, 8 January 1922. AHQP, microfilm AHQP/BSC-2, Section 1.

[31] Bohr to Ehrenfest, 19 May 1922. BCW2 1976, p. 631.

> This deals mostly with the general principles of quantum theory, and you will find that I have learned a lot from our discussions. You know how much the word expression means to me, and I can describe to you the situation not better than saying that I have felt that things themselves forced me to retrieve again the name Adiabatic Principle, and that I even have capitulated to the extent that I speak only about statistic weight, and had even fought against the use of a priori probability with the people here.

These remarks by Bohr appear in a different light if one reads the letter by Ehrenfest to which Bohr was responding.[32] It consists of a long and pathetic complaint regarding the contempt with which Sommerfeld had treated him in his latest edition of *Atombau*.[33] In it, Sommerfeld had reduced Ehrenfest's contribution to the adiabatic issue to coining the expression, and had attributed the fundamental and original idea to Einstein and Lorentz. Ehrenfest spent many pages explaining his own contributions, while acknowledging how Bohr had "formulated immediately in a very clear way which was the position of the degenerate systems," whereas he himself had "stood helpless before them."

Ehrenfest's participation in the subsequent developments of the adiabatic hypothesis after 1918 was limited. As far as publications are concerned, there is not a single paper devoted to the topic. He did mention the adiabatic hypothesis in a paper with Einstein in the summer of 1922, in which they discussed the experiment by Otto Stern and Walther Gerlach.[34] In this article they laid out the difficulties of understanding how the silver atoms become oriented as they travel across the varying magnetic field. One of the options considered entailed the elimination of the difference between slow and fast changes, a result they saw as problematic.

In June 1923, Ehrenfest himself had the opportunity to give his version of the birth of the adiabatic hypothesis in the commemorative issue of the Bohr atom with which we started our paper.[35] There, he reproduced in some more detail what he had written to Bohr the year before. In this 1923 paper, Ehrenfest emphasized that Bohr had formulated the most complete version of the adiabatic hypothesis and stressed the "organic" relation suggested by Bohr between the correspondence principle and the adiabatic principle in the new version of the theory, as well as his masterful treatment of degenerate systems.[36]

---

[32] Ehrenfest to Bohr, 8 May 1922. AHQP, microfilm AHQP/BSC-2, Section 1.

[33] Sommerfeld (1922), pp. 374-375.

[34] Einstein and Ehrenfest (1922).

[35] Ehrenfest (1923).
[36] Ehrenfest (1923), p. 543.

## 8. Final Remarks

Ehrenfest's adiabatic hypothesis, initially formulated by Ehrenfest in terms of (quantum-theoretically) allowed mechanical motions became, in the hands of Bohr, a necessary condition for the stability of the stationary states, as a logical extension of the sphere of validity of mechanics: only adiabatic transformations can be treated mechanically. This principle controlled the motions themselves as well as their statistical weights.

The adiabatic principle remained important in the later evolution of Bohr's theory. It turned out to have a relation to the correspondence principle, undoubtedly the more important principle after 1920. The masterful way in which Bohr tackled the degenerate motions Ehrenfest had bumped into, played a critical role in the establishment of this "organic" relation between the principles, as Ehrenfest pointed out in his paper of 1923. However, the calculations which could be performed with the help of the latter eclipsed the fundamental role of the adiabatic principle.

The later theorem of invariance of quantum numbers finally lost all connection to mechanics. Even then, Bohr never ceased to emphasize the crucial role of the adiabatic principle, making it widely known and reserving for it a special place in the history of the old quantum theory. Therefore, its role as a guiding principle in Bohr's efforts to increase the scope of the old quantum theory should not be underestimated.

*Acknowledgments*. We owe particular gratitude to Anthony Duncan and Michel Janssen for their helpful suggestions during the preparation of this paper. We also thank the editors and an anonymous referee for their critical comments, which have improved the early version of this paper.